\documentclass[a4paper,twocolumn]{jpsj3_ed}
\usepackage{times}
\usepackage{color}

\title{Density Functional Theory for Plasmon-assisted Superconductivity}

\author{$^{1}$Ryosuke Akashi\thanks{E-mail address: akashi@solis.t.u-tokyo.ac.jp} and
$^{1,2}$Ryotaro Arita
}
\inst{$^{1}$Department of Applied Physics, The University of Tokyo, Hongo, Bunkyo-ku, Tokyo 113-8656, Japan \\
$^{2}$JST-PRESTO, Kawaguchi, Saitama 332-0012, Japan
}
\abst{We review the recent progress in the density functional theory for superconductors (SCDFT). Motivated by the long-studied plasmon mechanism of superconductivity, we have constructed an exchange-correlation kernel entering the SCDFT gap equation which includes the plasmon effect. For the case of lithium under high pressures, we show that the plasmon effect substantially enhances the transition temperature ($T_{\rm c}$) by cooperating with the conventional phonon mechanism and results in a better agreement between the theoretical and experimentally observed $T_{\rm c}$. Our present formalism will be a first step to density functional theory for unconventional superconductors.}

\kword{density functional theory, superconductivity, phonon, plasmon, lithium, high pressure}

\begin{document}
\maketitle

\section{Introduction}
Superconductivity constitutes one of the most fascinating fields in condensed matter physics ever since its discovery in the early twentieth century. After the success of its description by the Bardeen-Cooper-Schrieffer theory,\cite{BCS} particular attention has been paid to the material dependence of the superconducting transition temperature ($T_{\rm c}$): that is, why some materials such as the celebrated cuprate\cite{Bednorz-Muller} exhibit high $T_{\rm c}$ but others do not? Since superconductivity emerges as a result of subtle interplay and competition of interactions between atoms and electrons having much larger energy scales, $T_{\rm c}$ is extremely sensitive to details of the electronic and crystal structure. Thus, an accurate quantitative treatment is essential to understand the emergence of high values of $T_{\rm c}$. 

For the conventional phonon-mediated mechanism, quantitative calculations have been performed within the Migdal-Eliashberg (ME) theory\cite{Migdal-Eliashberg} implemented with the first-principles method based on the Kohn-Sham density functional theory\cite{Kohn-Sham-eq}: In a variety of systems, phonon properties are well reproduced by the density functional perturbation theory\cite{Baroni-review} or the total-energy method\cite{Kunc-Martin-frozen} within the local density approximation\cite{Ceperley-Alder, PZ81}. By using the calculated phonon spectrum and electron-phonon coupling as inputs, it has been shown that the ME theory explains the qualitative tendency of $T_{\rm c}$ for various materials\cite{Savrasov-Savrasov, Choi-MgB2}. However, the ME formalism is not suitable for full \textit{ab initio} calculations since it is difficult to treat electron-electron interaction nonempirically. When we calculate $T_{\rm c}$ by solving the Eliashberg equation or using related approximate formulae such as the McMillan equation,\cite{McMillan,AllenDynes} we vary the value of $\mu^{\ast}$ (Ref.~\citen{Morel-Anderson}) representing the effective electron-electron Coulomb interaction which suppresses the Cooper-pair formation, and examine whether the range of the resulting $T_{\rm c}$ covers the experimentally observed value. With such a semi-empirical framework, the material dependence of the electron-electron interaction cannot be understood quantitatively. 

The recent progress in the density functional theory for superconductors (SCDFT)\cite{Oliveira, Kreibich, GrossI} has changed the situation. There, a non-empirical scheme describing the physics in the ME theory was formulated: Based on the Kohn-Sham orbital, it treats the weak-to-strong electron-phonon coupling, the screened electron-electron interaction within the static approximation, and the retardation effect\cite{Morel-Anderson} due to the difference in the energy ranges of these interactions. This scheme has been demonstrated to reproduce experimental $T_{\rm c}$s of various conventional phonon-mediated superconductors with deviation less than a few K.\cite{GrossII, Floris-MgB2, Sanna-CaC6, Bersier-CaBeSi} More recently, it has been employed to examine the validity of the ME theory in fully gapped superconductors with high $T_{\rm c}$ such as layered nitrides\cite{Akashi-MNCl} and alkali-doped fullerides.\cite{Akashi-fullerene} Through these applications, the current SCDFT has proved to be an informative method well-suited to investigate the nontrivial effects of the electron-electron interaction behind superconducting phenomena.

Although the electron-electron interaction just suppresses the pairing in the ME theory, possibilities of superconductivity induced by the electron-electron interaction have long been also explored. Since the discovery of the cuprates,\cite{Bednorz-Muller} superconductivity induced by short-range Coulomb interaction has been extensively investigated.\cite{Scalapino-review2012} On the other hand, there has been many proposals of superconducting mechanisms concerning long-range Coulomb interaction since the seminal work of Kohn and Luttinger.\cite{Kohn-Luttinger} In particular, there is a class of mechanisms that exploit the dynamical structure of the screened Coulomb interaction represented by the frequency-dependent dielectric function $\varepsilon(\omega)$: e.g., the plasmon\cite{Radhakrishnan1965,Frohlich1968,Takada1978,Rietschel-Sham1983} and exciton\cite{Little1967} mechanisms. Interestingly, such mechanisms can cooperate with the conventional phonon mechanism. Since they usually favor $s$-wave pairing, they have a chance to enhance $s$-wave superconductivity together with the phonon mechanism. Taking this possibility into account, these mechanisms are important even where they do not alone induce superconductivity. Therefore, they are expected to involve a broader range of systems than originally expected in the early studies. In fact, for a variety of systems having low-energy electronic excitations, theoretical model calculations addressing such a cooperation have been performed: SrTiO$_{3}$~\cite{Koonce-Cohen1967, Takada-SrTiO3} with small plasmon frequencies due to small electron densities, $s$-$d$ transition metals~\cite{Garland-sd} where ``demon" acoustic plasmons have been discussed,\cite{Pines-demon,Ihm-Cohen1981} metals sandwiched by small-gap semiconductors~\cite{Ginzburg-HTSC,ABB1973}, and layered systems where two-dimensional acoustic plasmons are proposed to become relevant~\cite{Kresin1987, Bill2002-2003}. Moreover, recent experimental discoveries of high-temperature superconductivity in doped band insulators have stimulated more quantitative analyses on effects of the cooperation~\cite{Yamanaka1998,Bill2002-2003, Taguchi2006, Taniguchi2012,Ye2012}.

Considering the above grounds, the situation calls for an \textit{ab initio} theory that treats the phonon-mediated interaction and the dynamical screened Coulomb interaction together, with which one can study from the superconductors governed by phonons or the dynamical Coulomb interaction to those by their cooperation on equal footing. The aim of our present study is to establish this by extending the applicability of SCDFT. In this paper, we review the recent theoretical extension to include the plasmon-induced dynamical screened Coulomb interaction.\cite{Akashi-plasmon} In Sec.~\ref{sec:theory}, we present the theoretical formulation and its practical implementation, and discuss how plasmons can enhance superconductivity. Section \ref{sec:appl-Li} describes the application to elemental lithium under high pressures, for which the plasmon effect is expected to be substantial because of its relatively dilute electron density. In Sec.~\ref{sec:summary} we summarize our results and give concluding remarks.

\section{Formulation}\label{sec:theory}
\subsection{General formalism}\label{subsec:theory-general}
Let us start from a brief review of SCDFT.\cite{GrossI} The current SCDFT employs the gap equation
\begin{eqnarray}
\Delta_{n{\bf k}}\!=\!-\mathcal{Z}_{n\!{\bf k}}\!\Delta_{n\!{\bf k}}
\!-\!\frac{1}{2}\!\sum_{n'\!{\bf k'}}\!\mathcal{K}_{n\!{\bf k}\!n'{\bf k}'}
\!\frac{\mathrm{tanh}[(\!\beta/2\!)\!E_{n'{\bf k'}}\!]}{E_{n'{\bf k'}}}\!\Delta_{n'\!{\bf k'}}
\label{eq:gap-eq}
\end{eqnarray}
to obtain $T_{\rm c}$, which is specified as the temperature where the calculated value of gap function $\Delta_{n{\bf k}}$ becomes zero. Here, $n$ and ${\bf k}$ denote the band index and crystal momentum, respectively, $\Delta$ is the gap function, and $\beta$ is the inverse temperature. The energy $E_{n {\bf k}}$ is defined as $E_{n {\bf k}}$=$\sqrt{\xi_{n {\bf k}}^{2}+\Delta_{n {\bf k}}^{2}}$ and $\xi_{n {\bf k}}=\epsilon_{n {\bf k}}-\mu$ is the one-electron energy measured from the chemical potential $\mu$, where $\epsilon_{n {\bf k}}$ is obtained by solving the normal Kohn-Sham equation in density functional theory
$
\mathcal{H}_{\rm KS}|\varphi_{n{\bf k}}\rangle=\epsilon_{n{\bf k}}
|\varphi_{n{\bf k}}\rangle
$
with $\mathcal{H}_{\rm KS}$ and $|\varphi_{n{\bf k}}\rangle$ being the Kohn-Sham Hamiltonian and the Kohn-Sham state, respectively. The functions $\mathcal{Z}$ and $\mathcal{K}$, which are called as the exchange-correlation kernels, describe the effects of all the interactions involved: They are defined as the second functional derivative of the free energy with respect to the anomalous electron density. A formulation of the free energy based on the Kohn-Sham perturbation theory\cite{} enables us practical calculations of the exchange-correlation functionals using the Kohn-Sham eigenvalues and eigenfunctions derived from standard \textit{ab initio} methods.

The nondiagonal exchange-correlation kernel $\mathcal{K}$ is composed of two parts $\mathcal{K}$$=$$\mathcal{K}^{\rm ph}$$+$$\mathcal{K}^{\rm el}$ representing the electron-phonon and electron-electron interactions, whereas the diagonal kernel $\mathcal{Z}$ consists of one contribution $\mathcal{Z}$$=$$\mathcal{Z}^{\rm ph}$ representing the mass renormalization of the normal-state band structure due to the electron-phonon coupling. The phonon parts, $\mathcal{K}^{\rm ph}$ and $\mathcal{Z}^{\rm ph}$, properly treats the conventional strong-coupling superconductivity. The electron-electron coutribution $\mathcal{K}^{\rm el}$ is the matrix element of the \textit{static} screened Coulomb interaction $\langle \varphi_{n{\bf k}\uparrow}\varphi_{n-{\bf k}\downarrow}|\varepsilon^{-1}(0)V|\varphi_{n'{\bf k}'\uparrow}\varphi_{n'-{\bf k}'\downarrow}\rangle$ with $V$ being the bare Coulomb interaction. Currently, the Thomas-Fermi approximation and the random-phase approximation (RPA) have been applied for the static dielectric function $\varepsilon^{-1}(0)$.\cite{Massidda} With these settings, the two parts of the nondiagonal kernel have different Kohn-Sham energy dependence: $\mathcal{K}^{\rm ph}$ has large values only for the states within the phonon energy scale, whereas $\mathcal{K}^{\rm el}$ decays slowly with the electronic energy scale. With this Kohn-Sham-state dependence, the retardation effect\cite{Morel-Anderson} is quantitatively treated. Thus, within the framework of the density functional theory, the SCDFT accurately treats the physics of Migdal-Eliashberg theory (based on the Green's function).

\begin{figure}[t]
 \begin{center}
  \includegraphics[scale=.16]{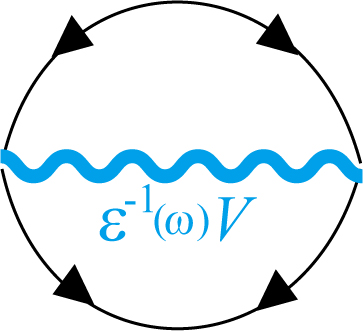}
  \caption{(Color online) Diagram corresponding to the electron nondiagonal kernel, $\mathcal{K}^{\rm el}$. The solid line with arrows running in the opposite direction denotes the electronic anomalous propagator~\cite{GrossI}. The blue wavy line denotes the screened electronic Coulomb interaction, which is a product of the inverse dielectric function $\varepsilon^{-1}$ and the bare Coulomb interaction $V$.}
  \label{fig:diagram}
 \end{center}
\end{figure}

The current setting $\mathcal{K}^{\rm el}$ $=$ $\langle \varphi_{n{\bf k}\uparrow}\varphi_{n-{\bf k}\downarrow}|$ $\varepsilon^{-1}(0)V$ $|\varphi_{n'{\bf k}'\uparrow}\varphi_{n'-{\bf k}'\downarrow}\rangle$ corresponds to the anomalous exchange contribution from the screened Coulomb interaction represented in Fig.~\ref{fig:diagram} with the $\omega$ dependence of $\varepsilon$ omitted. To incorporate effects of the plasmon on the interaction, we retain its frequency dependence. The diagram thus yields a following form 
\begin{eqnarray}
\hspace{-30pt}&&
\mathcal{K}^{\rm el, dyn}_{n{\bf k},n'{\bf k}}
\!=\!
\lim_{\{\Delta_{n{\bf k}}\}\rightarrow 0}
\frac{1}{{\rm tanh}[(\beta /2 ) E_{n{\bf k}}]}
\frac{1}{{\rm tanh}[(\beta /2) E_{n'{\bf k}'}]}
\nonumber \\
\hspace{-10pt}&&
\hspace{10pt}\times
\frac{1}{\beta^{2}}
\sum_{\tilde{\omega}_{1}\tilde{\omega}_{2}}
F_{n{\bf k}}({\rm i}\tilde{\omega}_{1})
F_{n'{\bf k}'}({\rm i}\tilde{\omega}_{2})
W_{n{\bf k}n'{\bf k}'}[{\rm i}(\tilde{\omega}_{1}\!\!-\!\!\tilde{\omega}_{2})]
,
\label{eq:kernel-dyn}
\end{eqnarray} 
where $F_{n{\bf k}}({\rm i}\tilde{\omega})$
$=$
$\frac{1}{{\rm i}\tilde{\omega}\!+\!E_{n{\bf k}}}
\!-\!
\frac{1}{{\rm i}\tilde{\omega}\!-\!E_{n{\bf k}}}
$
 and $\tilde{\omega}_{1}$ and $\tilde{\omega}_{2}$ denote the Fermionic Matsubara frequency. Function $W_{n{\bf k}n'{\bf k}'}({\rm i}\omega)$$\equiv$$\langle \varphi_{n{\bf k}\uparrow}\varphi_{n-{\bf k}\downarrow}|\varepsilon^{-1}({\rm i}\omega)V|\varphi_{n'{\bf k}'\uparrow}\varphi_{n'-{\bf k}'\downarrow}\rangle$ is the screened Coulomb interaction. We then apply the RPA\cite{RPA} to the $\omega$-dependent dielectric function, which is a standard approximation to describe the plasmon under crystal field. Formally, the present RPA kernel can be also derived from the RPA free energy defined by Eq.~(13) in Ref.~\citen{Kurth-Gross}: The set of terms of order $O(FF^{\dagger})$ (i.e., the set of the diagrams having only one anomalous bubble taken from Fig. 2 in Ref.~\citen{Kurth-Gross}) corresponds to the present kernel.

The Coulomb interaction $W_{n{\bf k}n'{\bf k}'}({\rm i}\nu)$ is practically calculated using a certain set of basis functions. Let us here summarize the plane-wave representation, which has been employed in our studies:
\begin{eqnarray}
&&\hspace{-30pt}W_{n{\bf k}n'{\bf k}'}({\rm i}\nu)
\nonumber \\
&& =
\frac{4\pi}{\Omega}\!
\sum_{{\bf G}\!{\bf G}'}\! 
\frac{
\!\rho^{n{\bf k}}_{n'{\bf k}'}(\!{\bf G}\!)\tilde{\varepsilon}^{-1}_{{\bf G}{\bf G}'}(\!{\bf k}\!-\!{\bf k}'\!;{\rm i}\nu)\!\{\rho^{n{\bf k}}_{n'{\bf k}'}(\!{\bf G}'\!)\!\}^*
}
{
|{\bf k}-{\bf k}'+{\bf G}||{\bf k}-{\bf k}'+{\bf G}'|
}\!,
\label{eq:K-el-RPA}
\end{eqnarray}
with $\tilde{\varepsilon}_{{\bf G}{\bf G}'}(\!{\bf k}\!-\!{\bf k}'\!;{\rm i}\nu)$ being the symmetrized dielectric matrix,\cite{Hybertsen-Louie} defined by
\begin{eqnarray}
\tilde{\varepsilon}_{{\bf G}{\bf G}'}({\bf K}; {\rm i}\nu)
\!\!\!\!&=&\!\!\!\!\!\!
\delta_{{\bf G}{\bf G}'}
\nonumber \\
&&\!\!\!-4\pi\frac{1}{|{\bf K}\!+\!{\bf G}|}\chi^{0}_{{\bf G}{\bf G}'}({\bf K}; {\rm i}\nu)\frac{1}{|{\bf K}\!+\!{\bf G}'|}
.
\end{eqnarray}
The independent-particle polarization $\chi^{0}_{{\bf G}{\bf G}'}({\bf K}; {\rm i}\nu)$ denotes 
\begin{eqnarray}
\chi^{0}_{{\bf G}{\bf G}'}({\bf K};{\rm i}\nu)
&\!\!\!\!=&\!\!\!\!
\frac{2}{\Omega}
\sum_{{\bf k}}\sum_{\substack{n:{\rm unocc}\\n':{\rm occ}}}
[\rho^{n{\bf k}+{\bf K}}_{n'{\bf k}}({\bf G})]^{\ast}\rho^{n{\bf k}+{\bf K}}_{n'{\bf k}}({\bf G}')
\nonumber \\
&& \hspace{-35pt}\times
[\frac{1}{{\rm i}\nu \!-\! \epsilon_{n {\bf k}+{\bf K}} \!+\! \epsilon_{n' {\bf k}}}
-
\frac{1}{{\rm i}\nu \!+\! \epsilon_{n {\bf k}+{\bf K}} \!-\! \epsilon_{n' {\bf k}}}]
,
\label{eq:chi-def}
\end{eqnarray}
where the band indices $n$ and $n'$ run through the unoccupied bands and occupied bands for each \textbf{k}, respectively. The matrix $\rho^{n'{\bf k}'}_{n{\bf k}}({\bf G})$ is defined by
\begin{eqnarray}
\rho^{n'{\bf k}'}_{n{\bf k}}({\bf G})
&=&
\int_{\Omega} d{\bf r}
\varphi^{\ast}_{n'{\bf k}'}({\bf r})
e^{{\rm i}({\bf k}'-{\bf k}+{\bf G})\cdot{\bf r}}
\varphi_{n{\bf k}}({\bf r}).
\label{eq:rho}
\end{eqnarray}
So far, we have ignored the intraband (Drude) contribution to $\tilde{\varepsilon}$ for ${\bf k}-{\bf k}'=0$: The kernel including this contribution diverges as $({\bf k}-{\bf k}')^{-2}$, whereas the total contribution by the small ${\bf k}-{\bf k}'$ to $T_{\rm c}$ should scale as $({\bf k}-{\bf k}')^{1}$ because of the ${\bf k}'$ integration in Eq.~(\ref{eq:gap-eq}).

\begin{figure}[b]
 \begin{center}
  \includegraphics[scale=.55]{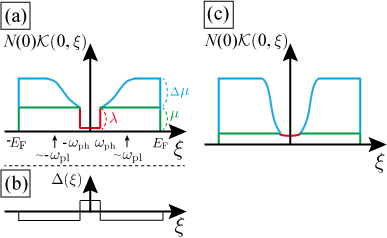}
  \caption{(Color online) (a) Energy dependence of nondiagonal kernels entering the gap equation. Phonon-induced attraction, static Coulomb repulsion, and the plasmon-induced high-energy Coulomb repulsion are indicated in red, green, and blue, respectively. (b) Approximate solution of the gap equation solved with the phonon and static Coulomb parts. (c) Energy dependence of the kernels in a case where the phonon part is negligibly small and the plasmon part is dominant.}
  \label{fig:interaction}
 \end{center}
\end{figure}

The physical meaning of the present dynamical correction to the previous static kernel is as follows. In real systems, screening by charge fluctuations is ineffective for the interaction with large energy exchanges [i.e., $\varepsilon(\omega) \xrightarrow{\omega \rightarrow \infty} 1$], whereas it becomes significant as the energy exchange becomes small compared with typical energies of charge excitations. However, the conventional static approximation ignores this energy dependence of the screening by extrapolating the static value of the interaction to the high energy, and underestimates the screened Coulomb repulsion with large energy exchanges. The present extension corrects this underestimation, and gives additional repulsive contribution to the Coulomb matrix elements between the Cooper pairs having much different energies. 

Interestingly, this additional contribution can raise $T_{\rm c}$. Let us discuss this point in terms of the interaction kernel entering the energy-averaged gap equation
\begin{eqnarray}
\Delta(\xi)
=
-\frac{1}{2}N(0)
\int \!\! d\xi' \!
\mathcal{K}(\xi\!,\xi')\frac{{\rm tanh}[(\beta/2)\xi']}{\xi'}\Delta(\xi')
,
\label{eq:gap-eq-ave}
\end{eqnarray}
where we define the averaged nondiagonal kernel as $\mathcal{K}(\xi,\xi')=\frac{1}{N(0)^{2}}\sum_{n{\bf k}n'{\bf k}'}\delta(\xi-\xi_{n{\bf k}})\delta(\xi'-\xi_{n'{\bf k}'})K_{n{\bf k}n'{\bf k}'}$ with $N(0)$ being electronic density of states at the Fermi level and omit the diagonal kernel for simplicity. This equation qualitatively describes coherent Cooper pairs represented by $\Delta(\xi)$ scattered by the pairing interactions. Suppose $\mathcal{K}=\mathcal{K}^{\rm ph}+\mathcal{K}^{\rm el}$, $N(0)\mathcal{K}^{\rm ph}(\xi,\xi')=-\lambda$ within the Debye frequency $\omega_{\rm ph}$ and $N(0)\mathcal{K}^{\rm el}(\xi,\xi')=\mu$ within a certain electronic energy range such as $E_{\rm F}$ (considering red and green parts in panel (a) of Fig.~\ref{fig:interaction}). Solving this equation by assuming $\Delta(\xi)$ to be nonzero and constant only within $\omega_{\rm ph}$, we obtain the BCS-type $T_{\rm c}$ formula $T_{\rm c}\propto \omega_{\rm ph}$$\times$$ {\rm exp}[-1/(\lambda-\mu)]$ for $\mu-\lambda<0$. However, if we allow $\Delta(\xi)$ to have nonzero constant values for $|\xi|>\omega_{\rm ph}$, we instead obtain $T_{\rm c}\propto \omega_{\rm ph}$$\times$${\rm exp}[-1/(\lambda-\mu^{\ast})]$ with $\mu^{\ast}=\mu/(1+\mu{\rm ln}[E_{\rm F}/\omega_{\rm ph}])<\mu$, and then, the resulting values of $\Delta(\xi)$ have opposite signs for $|\xi|<\omega_{\rm ph}$ and $|\xi|>\omega_{\rm ph}$ [panel (b) in Fig.~\ref{fig:interaction}]. Here, even if the total low-energy interaction $\mu-\lambda$ is repulsive, superconducting state realizes if $\mu^{\ast}-\lambda<0$. This weakening of the effective Coulomb repulsion is the celebrated retardation effect,\cite{Morel-Anderson} and its origin is the negative values of the high-energy gap function: Since the scattering by repulsion between Cooper pairs having $\Delta$ with opposite signs is equivalent to the scattering by attraction between those with same signs, there is a gain of the condensation energy.\cite{Kondo-PTP1963} Next, let us add the plasmon contribution [blue part in panel (a)], which enhances the repulsion by $\Delta \mu$ for $\xi$ with an energy scale of plasmon frequency $\omega_{\rm pl}$. Then, more condensation energy can be gained by enhancing the high-energy negative gap function, which increases $T_{\rm c}$. As an extreme situation, one can also consider the case where the phonon-induced attraction is negligible and the plasmon-induced repulsion is dominant [panel (c)]. Obviously, a superconducting solution exists even in this case because the discussion about the above $T_{\rm c}$ formula is also valid with the transformation $\lambda$$\rightarrow$$\Delta \mu$, $\mu$$\rightarrow$$\mu+\Delta \mu$ and $\omega_{\rm ph}$$\rightarrow$$\omega_{\rm pl}$. These discussions illustrate that the plasmon contribution can increase $T_{\rm c}$ by enhancing the high-energy repulsion. 

To the authors' knowledge, the plasmon mechanism of the above-mentioned type to enhance $T_{\rm c}$ has been originally studied by Takada\cite{Takada1978} based on the Green's function formalism for two- and three-dimensional homogeneous electron gas. Using the gap equation derived by himself, he has also performed calculations of $T_{\rm c}$ considering both the phonons and plasmons for doped SrTiO$_{3}$ (Ref.~\citen{Takada-SrTiO3}) and metal-intercalated graphites.\cite{Takada-graphite1982, Takada-graphite2009} Our present formalism, which treats the local field effect of inhomogeneous electron distribution behind the phonon and plasmon, is a DFT-based counterpart of his theory.\cite{comment-counterpart}

\subsection{Multipole plasmon approximation}\label{subsec:plasmon-pole}
Next we present a formulation to calculate $T_{\rm c}$ using the extended kernel. Evaluation of Eq.~(\ref{eq:kernel-dyn}) requires to perform the double discrete Matsubara summations for electronic energy scale, which is impractically demanding. We then analytically carry out the summations by approximating $W_{n{\bf k}n'{\bf k'}}$ as a simple function. For this purpose, we employ a multipole plasmon approximation
\begin{eqnarray}
\tilde{W}_{n{\bf k}n'{\bf k}'}({\rm i}\tilde{\nu}_{m})
\!\!\!\!&=&\!\!\!\!
W_{n{\bf k}n'{\bf k}'}(0)
\nonumber \\
&&+
\sum^{N_{\rm p}}_{i}
a_{i;n{\bf k}n'{\bf k}'}
g_{i;n{\bf k}n'{\bf k}'}(\tilde{\nu}_{m})
,
\label{eq:W-tilde}
\end{eqnarray}
with $g_{i;n{\bf k}n'{\bf k}'}$ being
\begin{eqnarray}
g_{i;n{\bf k}n'{\bf k}'}(x)
=
\frac{2}{\omega_{i;n{\bf k}n'{\bf k}'}}
-\frac{2\omega_{i;n{\bf k}n'{\bf k}'}}{x^{2}\!+\!\omega^{2}_{i;n{\bf k}n'{\bf k}'}}
.
\end{eqnarray}
Here, $\tilde{\nu}_{m}$ denotes the Bosonic Matsubara frequency. In contrast with the case of uniform electron gas, inhomogeneous systems can have a variety of plasmon modes, and our aim is to treat these modes in a unified manner. Substituting Eq.~(\ref{eq:W-tilde}) in Eq.~(\ref{eq:kernel-dyn}), we finally obtain $\mathcal{K}^{\rm el,dyn}$$=$$\mathcal{K}^{\rm el,stat}$$+$$\Delta\mathcal{K}^{\rm el}$ with $\mathcal{K}^{\rm el,stat}_{n{\bf k}n'{\bf k}'}$$=$$W_{n{\bf k}n'{\bf k}'}(0)$ and
\begin{eqnarray}
\hspace{-10pt}
\Delta\mathcal{K}^{\rm el}_{n{\bf k},n'{\bf k}}
&\!\!\!\!\!\!\!=&\!\!\!\!\!\!
\sum_{i}^{N_{\rm p}}\!2a_{i;n{\bf k}n'{\bf k}'} \!\left[
\frac{1}
{\omega_{i;n{\bf k}n'{\bf k}'}} \right.
\nonumber \\
&&
\hspace{-50pt}
\left.
+
\frac{
I\!(\xi_{n{\bf k}}\!,\!\xi_{n'{\bf k}'}\!,\omega_{i;n{\bf k}n'{\bf k}'}\!)
\!\!-\!\!
I\!(\xi_{n{\bf k}}\!,-\!\xi_{n'{\bf k}'}\!,\omega_{i;n{\bf k}n'{\bf k}'}\!)
}{{\rm tanh}[(\beta/2) \xi_{n{\bf k}}]{\rm tanh}[(\beta/2) \xi_{n'{\bf k}'}]}
\right]
,
\label{eq:Delta-kernel}
\end{eqnarray}
where the function $I$ is defined by Eq.~(55) in Ref.~\citen{GrossI}.

In order to calculate Eq.~(\ref{eq:Delta-kernel}), we determine the plasmon coupling coefficients $a_{i;n{\bf k}n'{\bf k}'}$ and the plasmon frequencies $\omega_{i;n{\bf k}n'{\bf k}'}$ by the following procedure: (i) Calculate the screened Coulomb interaction for the {\it real} frequency grid $W_{n{\bf k}n'{\bf k}'}(\nu_{j}\!+\!{\rm i}\eta)$, where \{$\nu_{j}$\} ($j=1, 2, . . .N_{\omega}$) specifies the frequency grid on which the numerical calculation is performed, and $\eta$ is a small positive parameter, (ii) determine the plasmon frequencies $\{\omega_{i;n{\bf k}n'{\bf k}'}\}$ by the position of the peaks up to the $N_{\rm p}$-th largest in ${\rm Im}W_{n{\bf k}n'{\bf k}'}(\nu_{j}\!+\!{\rm i}\eta)$, (iii) calculate the screened Coulomb interaction for the {\it imaginary} frequency grid $W_{n{\bf k}n'{\bf k}'}({\rm i}\nu_{j})$, and (iv) using the calculated $W_{n{\bf k}n'{\bf k}'}({\rm i}\nu_{j})$, determine the plasmon coupling coefficients $\{a_{i;n{\bf k}n'{\bf k}'}\}$ via the least squares fitting by $\tilde{W}_{n{\bf k}n'{\bf k}'}({\rm i}\nu_{j})$. 

For the fitting, the variance to be minimized is defined as
\begin{eqnarray}
S_{n{\bf k}n'{\bf k}'}
\!\!\!\!&=&\!\!\!\!
\sum^{N_{\omega }}_{j}
\delta \omega_{j}\biggl[
W_{n{\bf k}n'{\bf k}'}({\rm i}\nu_{j})
-W_{n{\bf k}n'{\bf k}'}(0)
\nonumber \\
&&
-
\sum^{N_{\rm p}}_{i}
a_{i;n{\bf k}n'{\bf k}'}
g_{i;n{\bf k}n'{\bf k}'}(\nu_{j})
\biggr]^{2}
,
\end{eqnarray}
and we have introduced a weight $\delta \omega_{j}$ satisfying $\sum^{N_{\omega }}_{j}\delta \omega_{j}$$=$$1$.
With all the plasmon frequencies given, the extrema condition $\frac{\partial S}{\partial a_{i}}=0$ ($i=1$, . . . , $N_{\rm p}$) reads
\begin{eqnarray}
\begin{pmatrix}
a_{1}\\
a_{2}\\
\vdots
\end{pmatrix}
\!\!\!\!&=&\!\!\!\!
\begin{pmatrix}
V^{gg}_{11} & V^{gg}_{12} & \cdots \\
V^{gg}_{21} & V^{gg}_{22} & \cdots \\
\vdots      & \vdots      & \ddots
\end{pmatrix}^{-1}
\begin{pmatrix}
V^{Wg}_{1} \\
V^{Wg}_{2} \\
\vdots
\end{pmatrix}
.
\label{eq:fit-coeff}
\end{eqnarray}
Here, $V^{Wg}$ and $V^{gg}$ are defined by
\begin{eqnarray}
V^{Wg}_{i}
&\!\!\!\!\!=&\!\!\!\!\!
\sum_{j=1}^{N_{\omega}}
\delta\omega_{j}[W_{j}-W(0)]
g_{i}(\nu_{j})
,\\
V^{gg}_{ij}
&\!\!\!\!\!=&\!\!\!\!\!
\sum_{k=1}^{N_{\omega}}
\delta\omega_{k}
g_{i}(\nu_{k})
g_{j}(\nu_{k})
.
\end{eqnarray}
For arbitrary frequency grids, we define the weight as
\begin{eqnarray}
\delta\omega_{j}\propto
\left\{
\begin{array}{cl}
0 & (j=1, N_{w}) \\
(\nu_{j+1}\!-\!\nu_{j-1})p_{j} & (j\neq 1, N_{w}) \\
\end{array}
\right.
.
\end{eqnarray}
The factor $p_{j}$ is the weight for the variance function introduced for generality, and we set $p_{j}=1$ in Secs.~\ref{sec:theory} and \ref{sec:appl-Li}. When a negative plasmon coupling appears, we fix the corresponding coupling to zero, recalculate Eq.~(\ref{eq:fit-coeff}), and repeat this procedure until all the coupling coefficients becomes nonnegative so that the positive definiteness of the loss function is guaranteed.

\begin{figure}[t]
 \begin{center}
  \includegraphics[scale=.51]{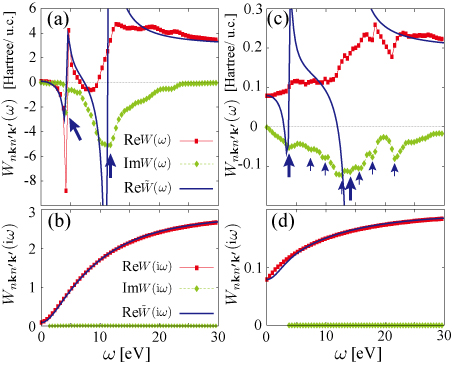}
  \caption{(Color online) Screened Coulomb interaction $W_{n{\bf k}n'{\bf k}'}$ and the corresponding approximate function $\tilde{W}_{n{\bf k}n'{\bf k}'}$ for fcc lithium under 14GPa calculated along the real frequency axis [(a), (c)], and the imaginary frequency axis [(b), (d)]. The band indices $n$ and $n'$ specify the partially occupied band. ${\bf k}$ and ${\bf k}'$ are $(2\pi/a)(1/7,1/7,1/7)$ and $(0,0,0)$ for (a)--(b), whereas $(2\pi/a)(2/7,2/7,6/7)$ and $(0,0,0)$ for (c)--(d).}
  \label{fig:fit}
 \end{center}
\end{figure}

For the determination of plasmon frequencies, the calculated spectrum of ${\rm Im}W_{n{\bf k}n'{\bf k}'}(\omega+{\rm i}\eta)$ is examined for each \{$n,{\bf k},n',{\bf k}'$\}. We have implemented a simple algorithm as follows: First, the peaks are specified as the point where the gradient of ${\rm Im}W_{n{\bf k}n'{\bf k}'}(\nu_{j}+{\rm i}\eta)$ turns from negative to positive; next, the specified peaks are sampled in order by their weighted values $p_{j}{\rm Im}W_{n{\bf k}n'{\bf k}'}(\nu_{j}+{\rm i}\eta)$. By increasing $N_{\rm p}$, we can expect all the relevant plasmon modes are properly considered.

We show in Fig.~\ref{fig:fit} the results of the fitting for fcc Li under 14GPa as typical cases where the fitting is straightforward [panel (a) and (b)] and difficult [(c) and (d)]. The peaks used for the fitting are indicated by arrows. For the former, an accurate fitting function was obtained with $N_{\rm p}$$=$$2$, where the derived fitting function and its analytic continuation $\tilde{W}_{n{\bf k}n'{\bf k}'}(\omega+{\rm i}\delta)$ indicated by thick blue lines reproduce the numerically calculated $W_{n{\bf k}n'{\bf k}'}({\rm i}\omega)$ and $W_{n{\bf k}n'{\bf k}'}(\omega+{\rm i}\delta)$ quite well, respectively. For the latter, on the other hand, good agreement of $\tilde{W}_{n{\bf k}n'{\bf k}'}({\rm i}\omega)$ and $W_{n{\bf k}n'{\bf k}'}({\rm i}\omega)$ was not achieved with $N_{\rm p}$$\leq$7, where $a_{i;n{\bf k}n'{\bf k}'}$ for the peaks indicated by the smaller arrows were zero. This was because one of the relevant plasmon modes indicated by the larger arrows was the eighth largest with respect to the peak height. The convergence of $T_{\rm c}$ with respect to $N_{\rm p}$ can be slow due to such a feature, though it becomes serious only for \{$n, {\bf k}, n', {\bf k}'$\} where the dynamical structure is blurred by strong plasmon damping [see the vertical axes in panels (a) and (c)]. 

We here also note possible systematic errors in the present algorithm. First, multiple plasmon peaks in $W_{n{\bf k}n'{\bf k}'}(\omega+{\rm i}\delta)$ may mutually overlap due to their peak broadening. Then, some plasmon modes are hidden by large broad peaks and cannot be specified even if we increase $N_{\rm p}$. We have assumed that these hidden modes are negligible because of their small spectral weight and strong damping. Next, the variance does not exactly converge to zero since the numerically calculated $W_{n{\bf k}n'{\bf k}'}({\rm i}\omega)$ shows a weak cusplike structure at ${\rm i}\omega=0$ [see panel (d) in Fig.~\ref{fig:fit}]. This structure probably originates from the finite lifetime of the plasmon modes. Its effect is not included by the plasmon-pole approximation, and will be examined in future studies.
\begin{table}[t!]
\caption[t]{Our calculated $T_{\rm c}$ considering only the phonon contributions (ph) to the exchange-correlation kernels, the phonon and static electron contributions (stat), and all the contributions ($N_{\rm p}$=1 and $N_{\rm p}$=2). Parameters $\lambda$$=$$2\int d\omega \alpha^{2}F(\omega)/\omega$ and $\omega_{\rm ln}$$=$${\rm exp}[\frac{2}{\lambda}\int d\omega {\rm ln}\omega\alpha^{2}F(\omega)/\omega]$ were calculated from the Eliashberg functions $\alpha^{2}\!F$ (Ref.~\citen{Migdal-Eliashberg}), and $r_{s}$$=$$\sqrt[3]{3/(4\pi \rho)}$ and $\Omega_{\rm p}$$=$$\sqrt{4\pi\rho/m^{\ast}}$ were calculated using the electron density $\rho$ and the band effective mass $m^{\ast}$, with $m^{\ast}$ evaluated from the fitting of the calculated density of states by that of the parabolic band.}
\begin{center}
\label{tab:Tc}
\begin{tabular}{lccccc}\hline
 & Al  &\multicolumn{4}{c}{fcc Li} \\
 &  &14GPa & 20GPa & 25GPa& 30GPa \\ \hline
$\lambda$ &0.417 &0.522 &0.623 &0.722 &0.812 \\
$\lambda^{a}$ &  & 0.49 &0.66 &  &0.83 \\
$\omega_{\rm ln}$ [K] & 314 &317&316&308&304 \\
$r_{s}$ &2.03 &2.71 &2.64  &2.59 &2.55 \\
$\Omega_{\rm p}$[eV] &16.2& 8.23& 8.44 &8.51 &8.58 \\ \hline
$T_{\rm c}^{\rm ph}$ [K]    &5.9 &10.0 &15.2 &19.0 &23.3  \\
$T_{\rm c}^{\rm stat}$ [K]  &0.8 &0.7  &1.8  & 3.2 &5.0           \\
$T_{\rm c}^{N_{\rm p}=1}$ [K]  &1.4 &2.2  &4.1 &6.5  &9.1          \\
$T_{\rm c}^{N_{\rm p}=2}$ [K]  &1.4 &2.2  &4.4 &6.8  &9.1 \\
$T_{\rm c}^{\rm expt.}$ [K]  & 1.20$^{b}$ & $<$4 &\multicolumn{3}{c}{5        --        17} \\ \hline
\multicolumn{6}{l}{$^{a}$\textit{Ab initio} density functional perturbation theory with the}\\
\multicolumn{6}{l}{Wannier-interpolation scheme, Ref.~\citen{Bazhirov-pressure}}\\
\multicolumn{6}{l}{$^{b}$ Ref.~\citen{Ashcroft-Mermin}}\\
\end{tabular} 
\end{center}
\end{table}
\begin{figure}[b!]
 \begin{center}
  \includegraphics[scale=.55]{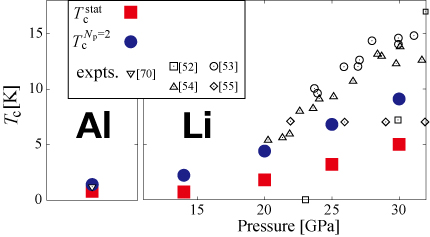}
  \caption{(Color online) Our calculated $T_{\rm c}$ (solid squares and circles) for aluminum and fcc lithium under high pressures compared with the experimentally observed values. The open symbols represent the experiments: Ref.~\citen{Ashcroft-Mermin} (open inverted triangle), Ref.~\citen{Shimizu2002} (open squares), Ref.~\citen{Struzhkin2002} (open circles), Ref.~\citen{Deemyad2003} (open regular triangles), and Ref.~\citen{Lin-Dunn} (open diamonds). }
  \label{fig:Tc-expt}
 \end{center}
\end{figure}
\section{Application to lithium under pressures}\label{sec:appl-Li}
The above formalism, which is based on the plasmon-pole approximation, is expected to be valid for the nearly uniform electron gas. We here present the recent application to an elemental-metal superconductor Li. Lithium has been known to exhibit superconductivity with $T_{\rm c}$$\gtrsim$10 K under high pressure.\cite{Shimizu2002,Struzhkin2002,Deemyad2003, Lin-Dunn} Early \textit{ab initio} calculations\cite{Christensen-Novikov, Tse, Kusakabe2005, Kasinathan, Jishi} including that based on the SCDFT\cite{Profeta-pressure} reproduced the experimentally observed pressure dependence of $T_{\rm c}$ quantitatively. However, a later sophisticated calculation\cite{Bazhirov-pressure} using the Wannier interpolation technique\cite{Giustino-Wannier-elph} has shown that the numerically converged electron-phonon coupling coefficient is far smaller than the previously reported values. On the other hand, the plasmon effect is expected to be substantial because the density of conducting electrons $n$, which determines a typical plasmon frequency by $\propto\sqrt{n}$, is relatively small in Li due to the large radius of the ion and the small number of valence electrons. Therefore, it is interesting to see if the newly included plasmon contribution fills the gap between the theory and experiment. It is also important to examine whether the present \textit{ab initio} method works successfully for conventional superconductors whose $T_{\rm c}$s have already been well reproduced by the conventional SCDFT. For that reason, we also applied the present method to aluminum.
\begin{figure*}[t!]
 \begin{center}
  \includegraphics[scale=.55]{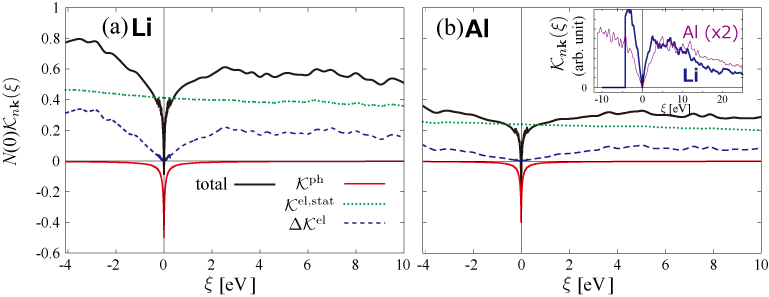}
  \caption{(Color online) Decomposition of the nondiagonal exchange-correlation kernel $\mathcal{K}_{n{\bf k}n'{\bf k}'}$ at $T$$=$$0.01$K calculated for (a) fcc lithium under pressure of 14GPa and (b) aluminum, averaged by equal-energy surfaces for $n'{\bf k}'$. The inset of panel (b) compares $\Delta\mathcal{K}^{\rm el}$ of lithium and aluminum, where the latter values are doubled to stress the difference in the energy scale. The abrupt onset in negative-energy region specifies the position of the conduction-band bottom.}
  \label{fig:kernel}
 \end{center}
\end{figure*}
\subsection{Calculation with small $N_{p}$}\label{subsec:small-Np}
In Ref.~\citen{Akashi-plasmon}, we performed calculations for fcc Li under pressures of 14, 20, 25, and 30GPa. All our calculations were carried out within the local-density approximation~\cite{Ceperley-Alder,PZ81} using {\it ab initio} plane-wave pseudopotential calculation codes {\sc Quantum Espresso}~\cite{Espresso,Troullier-Martins} (see Ref.~\citen{Akashi-plasmon} for further details). The phonon contributions to the SCDFT exchange-correlation kernels ($\mathcal{K}^{\rm ph}$ and $\mathcal{Z}^{\rm ph}$) were calculated using the energy-averaged approximation~\cite{GrossII}, whereas the electron contributions ($\mathcal{K}^{\rm el,stat}$ and $\Delta\mathcal{K}^{\rm el}$) were calculated by Eq.~(13) in Ref.~\citen{Massidda} and Eq.~(\ref{eq:Delta-kernel}) to evaluate the plasmon effect. The SCDFT gap equation was solved with a random sampling scheme given in Ref.~\citen{Akashi-MNCl}, with which the sampling error in the calculated $T_{\rm c}$ was not more than a few percent. In addition to the typical plasmon, an extra plasmon due to a band-structure effect has been discussed for Li\cite{Karlsson-Aryasetiawan, Silkin2007} and Al\cite{Hoo-Hopfield, Sturm-Oliveira1989}. We therefore carried out the calculation for $N_{\rm p}$$=$$1$ and $2$. 

In Table \ref{tab:Tc}, we summarize our calculated $T_{\rm c}$ values with $\mathcal{K}$$=$$\mathcal{K}^{\rm ph}$ ($T_{\rm c}^{\rm ph}$), $\mathcal{K}$$=$$\mathcal{K}^{\rm ph}$$+$$\mathcal{K}^{\rm el,stat}$ ($T_{\rm c}^{\rm stat}$), and $\mathcal{K}$$=$$\mathcal{K}^{\rm ph}$$+$$\mathcal{K}^{\rm el,stat}$$+$$\Delta\mathcal{K}^{\rm el}$ ($T_{\rm c}^{N_{\rm p}=1}$ and $T_{\rm c}^{N_{\rm p}=2}$). The estimated electron-phonon coupling coefficient $\lambda$, the logarithmic average of phonon frequencies $\omega_{\rm ln}$, the density parameter $r_{s}$, and typical plasma frequency $\Omega_{\rm p}$ are also given. Instead of using the Wannier-interpolation technique, we carried out the Fermi surface integration for the input Eliashberg functions\cite{Migdal-Eliashberg} with broad smearing functions,~\cite{Akashi-plasmon} and we obtained $\lambda$ consistent with the latest calculation~\cite{Bazhirov-pressure}, which is smaller than the earlier estimates~\cite{Tse,Kusakabe2005, Profeta-pressure,Kasinathan,Christensen-Novikov,Jishi}. The material and pressure dependence of the theoretical $T_{\rm c}$ follows that of $\lambda$. With $\mathcal{K}$$=$$\mathcal{K}^{\rm ph}$, $T_{\rm c}$ is estimated to be of order of 10K. While it is significantly suppressed by including $\mathcal{K}^{\rm el,stat}$, it is again increased by introducing $\Delta\mathcal{K}^{\rm el}$. We here do not see significant $N_{\rm p}$ dependence, which is further examined in Sec.~\ref{subsec:large-Np}.

The calculated values of $T_{\rm c}$ are compared with the experimental values in Fig.~\ref{fig:Tc-expt}. With the static approximation (red square), the general trend of the experimentally observed $T_{\rm c}$ is well reproduced: Aluminum exhibits the lowest $T_{\rm c}$, and $T_{\rm c}$ in Li increases as the pressure becomes higher.
However, the calculated $T_{\rm c}$ for Li is significantly lower than the experimental one, which demonstrates that the conventional phonon theory is quantitatively insufficient to understand the origin of the high $T_{\rm c}$ in Li under high pressures. From the previous \textit{ab initio} calculations, this insufficiency has not been well recognized because either too strong electron-phonon coupling or too weak electron-electron Coulomb interaction was used. With the plasmon contribution (blue circle), the resulting $T_{\rm c}$ systematically increases compared with the static-level one and it becomes quantitatively consistent with the experiment. For Al, in contrast, the accuracy is acceptable with both $T^{\rm stat}_{\rm c}$ and $T^{N_{\rm p}=2}_{\rm c}$, where the increase of $T_{\rm c}$ by $\Delta\mathcal{K}^{\rm el}$ is relatively small. These results indicate the followings: First, the plasmon contribution is essential for the high $T_{\rm c}$ in fcc Li under pressure, and second, our scheme gives accurate estimates of $T_{\rm c}$ regardless of whether their dynamical effects are strong or weak. 

\begin{figure}[t!]
 \begin{center}
  \includegraphics[scale=.6]{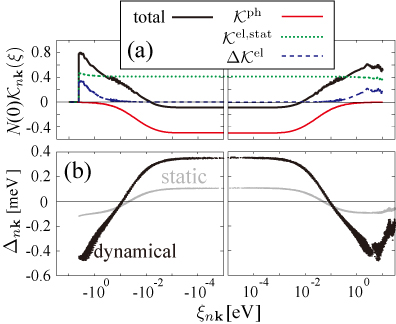}
  \caption{(Color online) (a) Decomposition of the nondiagonal exchange-correlation kernel $\mathcal{K}_{n{\bf k}n'{\bf k}'}$ at $T$$=$$0.01$K calculated for fcc lithium under pressure of 14GPa, averaged by equal-energy surfaces for $n'{\bf k}'$. (b) The corresponding gap function calculated with (darker) and without (lighter) $\Delta\mathcal{K}^{\rm el}$.}
  \label{fig:kernel-gap}
 \end{center}
\end{figure}

We discuss the origin of the enhancement of $T_{\rm c}$ by considering the dynamical effect in terms of partially energy-averaged nondiagonal kernels $\mathcal{K}_{n{\bf k}}(\xi)$$\equiv$$\frac{1}{N(\xi)}\sum_{n'{\bf k}'}\mathcal{K}_{n{\bf k}n'{\bf k}'}\delta(\xi-\xi_{n'{\bf k}'})$. With $n{\bf k}$ chosen as a certain point near the Fermi energy, we plotted the averaged kernel for fcc Li under pressure of 14GPa and Al with $N_{\rm p}$$=$$2$ in Fig.~\ref{fig:kernel}. The total kernel is decomposed into $\mathcal{K}^{\rm ph}$ (solid red line), $\mathcal{K}^{\rm el,stat}$ (dotted green line), and $\Delta\mathcal{K}^{\rm el}$ (dashed blue line). Generally, the total kernel becomes slightly negative within the energy scale of the phonons due to $\mathcal{K}^{\rm ph}$, whereas it becomes positive out of this energy scale mainly because of $\mathcal{K}^{\rm el,stat}$. The $\Delta\mathcal{K}^{\rm el}$ value is positive definite, but nearly zero for a low energy scale. As discussed in Sec.~\ref{sec:theory}, the high-energy enhancement of repulsion increases $T_{\rm c}$ through the retardation effect. Remarkably, $\Delta\mathcal{K}^{\rm el}$ sets in from an energy far smaller than the typical plasmon frequency (see Table \ref{tab:Tc}), and its absolute value is of the same order of $\mathcal{K}^{\rm el,stat}$. These features can be also seen in the case of homogeneous electron gas studied by Takada~\cite{Takada1978}. On the difference between Li and Al [(a) and (b)], we see that the contribution of $\Delta\mathcal{K}^{\rm el}$ in Al is noticeably smaller than that in Li. Also, the energy scale of the structure of $\Delta\mathcal{K}^{\rm el}$ [inset of (b)], which correlates with $\Omega_{\rm p}$ (see Table \ref{tab:Tc}), is small (large) for Li (Al). These differences explain why the effect of $\Delta\mathcal{K}^{\rm el}$ is more significant in Li.

The enhanced retardation effect by the plasmon is seen more clearly from the gap functions plotted together with the non-diagonal kernel in Fig.~\ref{fig:kernel-gap}. Indeed, we observe substantial enhancement of the negative gap value in the high-energy region, where the additional repulsion due to $\Delta\mathcal{K}^{\rm el}$ is strong. This clearly demonstrates that the plasmon mechanism indeed enhances $T_{\rm c}$, as is described in Sec.~\ref{sec:theory}.

We did not find a nonzero solution for the gap equation Eq.~(\ref{eq:gap-eq}) with only the electron-electron contributions ($\mathcal{K}$$=$$\mathcal{K}^{\rm el,stat}$$+$$\Delta\mathcal{K}^{\rm el}$) down to $T=0.01$ kelvin, but did with the electron-phonon and the static electron-electron contribution ($\mathcal{K}$$=$$\mathcal{K}^{\rm ph}$$+$$\mathcal{K}^{\rm el,stat}$). Hence, while the driving force of the superconducting transition in Li is the phonon effect, the plasmon effect is essential to realize high $T_{\rm c}$.

Finally, we also examined the effect of energy dependence of electronic density of states (DOS) on $\mathcal{Z}^{\rm ph}$. Since the form for $\mathcal{Z}^{\rm ph}$ used above [Eq.~(24) in Ref.~\citen{GrossII}] only treats the constant component of the density of states, we also employed a form generalized for the nonconstant density of states [Eqs.~(40) in Ref.~\citen{Akashi-asym}]. The calculated $T_{\rm c}$ changes by approximately 2\% with the nonconstant component, indicating that the constant-DOS approximation for the phonon contributions is valid for the present systems.

\begin{figure}[t]
\begin{center}
\includegraphics[width=7cm]{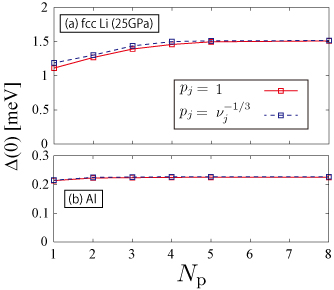}
\end{center}
\caption{(Color online) $N_{\rm p}$ dependence of the calculated gap function $\Delta_{n{\bf k}}$ at the Fermi level at $T$=0.01 K for (a) fcc lithium under pressure of 25GPa and (b) aluminum.}
\label{fig:gap-conv}
\end{figure}

\begin{table}[b]
\caption[t]{Calculated $T_{\rm c}$ with different $N_{\rm p}$ using the procedure described in the text. For $N_{\rm p}$$=$1 and 2, the calculated values in Table \ref{tab:Tc} are given together for comparison (left values).}
\begin{center}
\label{tab:Tc-2}
\begin{tabular}{lccccc}\hline
 & Al  &\multicolumn{4}{c}{fcc Li} \\
 &  &14GPa & 20GPa & 25GPa& 30GPa \\ \hline
$T_{\rm c}^{N_{\rm p}=1}$ [K]  &1.4,1.5 &2.2,2.8  &4.1,5.2 &6.5,7.4  &9.1,11.1          \\
$T_{\rm c}^{N_{\rm p}=2}$ [K]  &1.4,1.6 &2.2,3.1  &4.4,5.5 &6.8,8.0  &9.1,10.7 \\
$T_{\rm c}^{N_{\rm p}=5}$ [K]  &1.6 &3.8  &6.5 &9.2&12.0 \\
$T_{\rm c}^{N_{\rm p}=8}$ [K]  &1.6 &3.8  &6.5 &9.2&12.0 \\ \hline
\end{tabular} 
\end{center}
\end{table}

\subsection{$N_{\rm p}$ dependence of $T_{\rm c}$}\label{subsec:large-Np}
Here we investigate the convergence of $T_{\rm c}$ with respect to the number of plasmon peaks $N_{\rm p}$. To address this problem, on top of the procedure described in Secs. \ref{sec:theory} and \ref{subsec:small-Np}, we employed a slightly different algorithm. The difference is as follows. First, in the previous procedure, the plasmon frequencies $\omega_{i;n{\bf k}n'{\bf k}'}$ and coupling coefficients $a_{i;n{\bf k}n'{\bf k}'}$ for a set of sampling points were calculated from linear interpolation using the \textit{ab initio} data on the equal grid, where the interpolation was independently carried out for each $i$-th largest branch. Since such an algorithm becomes unstable for damped peaks, we here did not carry out that, but rather determined $\omega_{i;n{\bf k}n'{\bf k}'}$ and $a_{i;n{\bf k}n'{\bf k}'}$ simply by the \textit{ab initio} values on the neighboring grid point. Second, the weight for the variance and the ordering of the peak $p_{j}$ (see Sec.~\ref{subsec:plasmon-pole}) was set to unity in the previous procedure, but we here adopted $p_{j}= \nu_{j}^{-(1/3)}$: In an analytic $T_{\rm c}$ formula for three-dimensional electron gas derived by Takada [Eq.~(2.28) in Ref.~\citen{Takada1978}], the coefficient $\langle F \rangle$ in the exponent depends on the typical plasmon frequency by $\Omega_{\rm p}^{-(1/3)}$, so that we determined $p_{j}$ accordingly. We have indeed found that this setting of $p_{j}$ accelerates the convergence of the calculated gap function with respect to $N_{\rm p}$, as demonstrated by Fig.~\ref{fig:gap-conv}.\cite{comment-accelerate}

Carrying out the above procedure,\cite{comment-recalc} we calculated $T_{\rm c}$ for Al and Li under the pressures. The calculated result for $N_{\rm p}$$=$1, 2, 5 and 8 is summarized in Table \ref{tab:Tc-2} together with that of Sec.~\ref{subsec:small-Np}. For $N_{\rm p}$$=$1 and 2, the previous and present procedures give slightly different values of $T_{\rm c}$, which originates mainly from the difference in the interpolation of $\omega_{i;n{\bf k}n'{\bf k}'}$ and $a_{i;n{\bf k}n'{\bf k}'}$. Within the present results, the caluculated $T_{\rm c}$ for Al shows little $N_{\rm p}$ dependence, whereas $N_{\rm p}$ has to be larger than 5 for Li to achieve the convergence within 0.1K. This indicates that the damped dynamical structure of the Coulomb interaction ignored with $N_{\rm p}$$=$1 and 2 also has a nonnegligible effect. We note that the general numerical trend observed in the results in Sec.~\ref{subsec:small-Np} is also valid for the calculated values with $N_{\rm p}\geq 5$.

\section{Summary and Conclusion}\label{sec:summary}
We reviewed the recent progress by the authors in the SCDFT to address non-phonon superconducting mechanisms.\cite{Akashi-plasmon} An exchange-correlation kernel entering the SCDFT gap equation has been formulated within the dynamical RPA so that the plasmons in solids are considered. Through the retardation effect, plasmons can induce superconductivity, which has been studied for more than 35 years as the plasmon-induced pairing mechanism. A practical method to calculate $T_{\rm c}$ considering the plasmon effect have been implemented and applied to fcc Li. We have shown that the plasmon effect considerably raises $T_{\rm c}$ by cooperating with the conventional phonon-mediated pairing interaction, which is essential to understand the high $T_{\rm c}$ in Li under high pressures. 

The recent application suggests a general possibility that plasmons have a substantial effect on $T_{\rm c}$, even in cases where it does not alone induce superconducting transition. It is then interesting to apply the present formalism to ``other high-temperature superconductors"\cite{Pickett-review-other} such as layered nitrides, fullerides, and the bismuth perovskite. Effects of the electron-electron and electron-phonon interactions in these systems have recently been examined from various viewpoints, particularly with \textit{ab initio} calculations.\cite{Meregalli-Savrasov-BKBO, Heid-Bohnen2005,Yin-Kotliar-PRX,Antropov-Gunnarsson-C60,Janssen-Cohen-C60,Akashi-MNCl,Akashi-fullerene, Nomura-C60-cRPA} Since they have a nodeless superconducting gap, plasmons may play a crucial role to realize their high $T_{\rm c}$.\cite{Bill2002-2003} More generally, there can be other situations: (i) the phonon effect does not dominate over the static Coulomb repulsion, but the plasmon effect does (i.e., a superconducting solution is not found with $\mathcal{K}$$=$$\mathcal{K}^{\rm ph}$$+$$\mathcal{K}^{\rm el,stat}$, but is found with $\mathcal{K}$$=$$\mathcal{K}^{\rm el,stat}$$+$$\Delta\mathcal{K}^{\rm el}$), and (ii) either of the two effects does not independently, but their cooperation does (i.e., a superconducting solution is found with $\mathcal{K}$$=$$\mathcal{K}^{\rm ph}$$+$$\mathcal{K}^{\rm el,stat}$$+$$\Delta\mathcal{K}^{\rm el}$). Searching for superconducting systems of such kinds is another interesting future subject, for which our scheme provides a powerful tool based on the density functional theory.

\section*{Acknowledgments}
The authors thank Kazuma Nakamura and Yoshiro Nohara for providing subroutines for calculating the RPA dielectric functions. This work was supported by Funding Program for World-Leading Innovative R \& D on Science and Technology (FIRST Program) on ``Quantum Science on Strong Correlation,'' JST-PRESTO, Grants-in-Aid for Scientic Research (No. 23340095), and the Next Generation Super Computing Project and Nanoscience Program from MEXT, Japan.



\begin{thebibliography}{999}
\bibitem{comment-counterpart} His theory also treats the coupling between phonons and plasmons, which is not considered in our method.
\bibitem{comment-accelerate} The two settings of $p_{j}$ give slightly different values of the variance for $N_{\rm p}\rightarrow \infty$ due to the cusplike structure discussed in Sec.~\ref{subsec:plasmon-pole}, though the difference is invisibly small here.
\end{thebibliography}
\end{document}